\documentclass[fleqn,twoside]{article}
\usepackage{espcrc2}
\usepackage{graphicx}
\begin{document}
\title{QED in 2+1 Dimensions with Fermi and Gap Anisotropies}
\author{Iorwerth Owain Thomas and  Simon Hands\\
	{\em Department of Physics}, {\em University of Wales Swansea,} \\
	{\em Singleton Park, Swansea}, {\em United Kingdom SA2 8PP}}
\date{started 01/06/2004, last modified: \today}

\begin{abstract}
 QED in 2+1 dimensions has long been studied as a model field theory which exhibits both asymptotic freedom and non-trivial IR behaviour.  There is also a trend towards viewing it as a candidate low energy effective theory for the pseudogap phase of high temperature superconductors.  One feature of these theories is their lack of isotropy in the $x$ and $y$ directions (a common feature of Dirac theories in condensed matter systems).  A model motivated by this work is outlined, and tentative evidence presented for chiral symmetry restoration as the relative anisotropy is increased.
\vspace{1pc}
\end{abstract}
\maketitle

\section{Introduction}

One of the outstanding questions in the theory of condensed matter systems is the nature of the so-called {\em pseudogap} phase of the cuprate family of high temperature superconductors.  

This phase, which is reached from the superconducting (dSC) phase by a combination of increased temperature and reduced doping (x), is characterised by the persistence of an energy gap but without the corresponding sharp quasiparticle peak --- that is, an energy gap but no superconductivity.  Traditionally, approaches to this problem begin with a model describing the antiferromagnetic (AF) phase as a starting point and attempt to derive the properties of the pseudogap from it; however, a more recent trend, due to \cite{kn:Fishy},  begins from the dSC phase instead.  Inspired by this approach, the authors of \cite{kn:Tes0} and \cite{kn:Herb0} have derived a low temperature effective theory which very much resembles quantum electrodynamics in 2+1 dimensions (QED$_3$) and this in turn has motivated our research into the area.

\section{From dSC to QED$_3$: A brief review}

Following \cite{kn:Herb0}, we begin with the action of a d-wave superconductor at $T \neq 0$:
\begin{eqnarray}
\lefteqn{S=T\sum_{{\bf k}, \sigma, \omega_{n}}[(i\omega_{n}-\xi_{{\bf k}})c_{\sigma}^{\dag}({\bf k},\omega_{n})c_{\sigma}({\bf k},\omega_{n})} \nonumber\\
 &-& \frac{\sigma}{2}\Delta({\bf k})c_{\sigma}^{\dag}({\bf k},\omega_{n})c_{-\sigma}^{\dag}(-{\bf k},-\omega_{n}) \nonumber\\ 
&+& h.c. + O(c^{4})]			\label{eq:dSC}
\end{eqnarray}
where $c_{\sigma}^{\dag}$, $c_{\sigma}$ are the $e^{-}$ operators, $\sigma=+,-$ signifies spin, $\omega_{n}$ is the fermionic Matsubara frequency and $O(c^{4})$ are terms governing short range interactions.  $\Delta({\bf k})$ is the gap function, which has d-wave symmetry.

We expand (\ref{eq:dSC}) around the four nodes of the d-wave Fermi surface, and rewrite in terms of the fields 
\small{$\Psi_{i}^{\dag}({\bf q},\omega_{n}) = (c_{+}^{\dag}({\bf k},\omega_{n}),c_{-}(-{\bf k},-\omega_{n}),c_{+}^{\dag}({\bf k}- {\bf Q}_{i},\omega_{n}),c_{-\sigma}(-{\bf k}+{\bf Q}_{i},-\omega_{n}))$}. \normalsize Here, $i=1,2$ and the wave vectors linking the nodes into diagonal pairs are ${\bf Q}_{i}=2{\bf K}_{i}$, where ${\bf K}_{i}$ is the distance of the node from the origin in momentum space.

For \small ${\bf k}={\bf K}_{i}+{\bf q}$\normalsize, \small$|{\bf q}|\ll |{\bf K}_{i}|$ \normalsize we can linearise certain parameters as $ \xi_{{\bf k}}=v_{F}q_{x} + O(q^{2})$, $\Delta({\bf k})=v_{\Delta}q_{y} + O(q^{2})$.  $v_{F}$, $v_{\Delta}$ are referred to as the {\em Fermi} and {\em Gap} velocities respectively.  If we define \small$\delta=\sqrt{v_{F}v_{\Delta}}$ \normalsize  and the {\em anisotropy parameter}\footnote{This parameter depends on the doping x of the superconductor. \cite{kn:Herb1} states that the value of this parameter at optimum doping is about 10.} \small$\kappa=\frac{v_{F}}{v_{\Delta}}$\normalsize, as is done in \cite{kn:Herb1}, we can make the replacements \small $v_{F}p_{x} \rightarrow \frac{\delta}{\sqrt{\kappa}}p_{x}$ \normalsize and \small$v_{\Delta}p_{y} \rightarrow \delta\sqrt{\kappa}p_{x}$ \normalsize in the linearisation; this makes the comparison of results with the predictions within \cite{kn:Herb1} easier.  Transforming the result of our manipulations into position space we have:
\begin{eqnarray}
	\lefteqn{S= \int d^{2}{\bf r} \, d\tau \, \{\bar{\Psi}_{1}[\gamma_{0}\partial_{\tau}  + \frac{\delta}{\sqrt{\kappa}}\gamma_{1}\partial_{x}  + \delta\sqrt{\kappa}\gamma_{2}\partial_{y}]\Psi_{1}} \nonumber \\
  	&+& (1 \rightarrow 2, x \leftrightarrow y)\} 	\label{eq:QEDnogauge}
\end{eqnarray}
This describes a noninteracting system; the phase degree of freedom in the order parameter $\Delta({\bf k})$ allows us to add gauge fields and interaction terms.  Detailed workings out of this are given in \cite{kn:Tes0,kn:Herb0}, but for our purposes, the more heurisitic accounts given in \cite{kn:Vort0,kn:Vort1} will suffice.  They proceed as follows: the phase d.o.f. allows the formation of vortices.  These vortices, if they are allowed to condense, will disorder the dSC phase, placing us within the pseudogap.  The photon in 2+1 dimensions has the same physical d.o.f. as the Goldstone boson produced by vortex condensation, so it is natural to model the effects of the phase vortices as a minimally coupled U(1) gauge theory\footnote{The correct approach for modelling this on the lattice is to use noncompact formulations of U(1).  Due to instanton formation, compact formulations do not preserve flux symmetry ($\partial_{\mu}j_{\mu}\neq0$), hence the photon in such a model is {\em massive} \cite{kn:Vort0}, \cite{kn:Vort1}.}.  Doing this, we can write the anisotropic QED$_3$ action of \cite{kn:Tes0,kn:Herb0,kn:Herb1}, where $a_{\mu}$ is the photon field, and $e$ is expectation value of  the vortex condensate $\langle\Phi\rangle$:
\begin{eqnarray}
	\lefteqn{S= \int d^{2}{\bf r} \, d\tau \, \{\bar{\Psi}_{1}[\gamma_{0}(\partial_{\tau} +ia_{\tau}) + \frac{\delta}{\sqrt{\kappa}}\gamma_{1}(\partial_{x} +ia_{x})} \nonumber\\
&+& \delta\sqrt{\kappa}\gamma_{2}(\partial_{y} +ia_{y})]\Psi_{1}+ (1 \rightarrow 2, x \leftrightarrow y) \nonumber\\
&+& \frac{1}{4e^{2}}F_{\mu\nu}F_{\mu\nu}\} \label{eqn:fullQED3}
\end{eqnarray}

\section{Relevance of chiral symmetry breaking}
One of the more interesting questions regarding the phase diagram at $T=0$ is whether the pseudogap exists as an intermediate phase between the AF phase and the dSC phase.  \cite{kn:Herb0} and \cite{kn:Tes0} show that the chiral condensate $\langle\bar{\Psi}\Psi\rangle$ can be identified with a Spin Density Wave order parameter (which becomes a full AF phase as x decreases).  Consequently, $\langle\bar{\Psi}\Psi\rangle=0$ signals the existence of the pseudogap.  Note that at no point can our model enter the dSC phase; the addition of the gauge fields to our action presupposed that we had left it.  

The much vexed question of the critical number of flavours ($N_{c}$) required for chiral symmetry breaking in QED$_{3}$ now becomes relevant.  If $N_{c}<2$ $\forall\kappa$, then no pseudogap exists as an intermediate phase at $T=0$.  If $N_{c}>2$ $\forall\kappa$, then it does.  Schwinger-Dyson techniques have given values for $N_{c}$ of $\infty$, $\frac{32}{\pi^{2}}$ and $\sim4$ (see references cited in \cite{kn:Hands0}).  A thermodynamic argument due to \cite{kn:Appel0} has $N_{c}<\frac{3}{2}$, whilst \cite{kn:Hands0} has an argument that $N_{c}>4$ due to a correspondence with the Thirring model.  Lattice calculations \cite{kn:Hands0} have failed to shed much light on the issue; chiral condensates (after the appropriate limits are taken, and allowing for the finite size effects that bedevil QED$_3$ simulations) are very small, which indicates either symmetry restoration, or symmetry breaking with a very small condensate.  It is hoped that despite these issues, simulations of the anisotropic case might allow general conclusions to be drawn as to the nature of the phase diagram.

\section{The Lattice Model}

\begin{figure}[t]
	\input{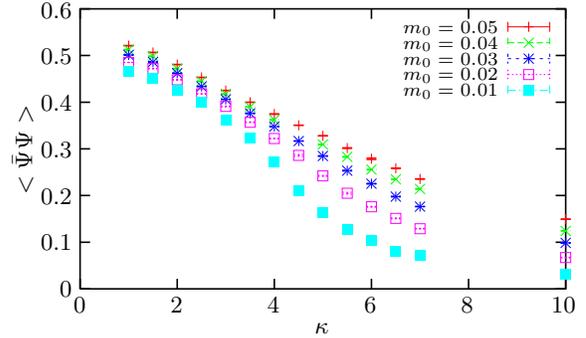}
\vspace{-\textfloatsep}
	\caption{$\langle\bar{\Psi}\Psi\rangle$ versus $\kappa$ for $m_{0}$ between 0.01 and 0.05.} \label{fig:chircon}
\vspace{-.75\textfloatsep}
\end{figure}
We simulate the following staggered fermion action using a Hybrid Monte-Carlo algorithm on a 16$^{3}$ lattice:
\begin{equation}
	S =\sum_{i=1}^{N} \sum_{x,x'} \bar{\chi_{i}}(x) M_{x,x'} \chi_{i}(x') 
 +\frac{\beta}{2} \!\!\sum_{x,\mu<\nu}\!\! \Theta_{\mu\nu}^{2}(x) \label{eqn:lattact}
\end{equation}

 The fermion matrix is defined as follows:
\begin{equation}
	M_{x,x'} =  \sum_{\mu=1}^{3} \xi_{\mu}(x) [\delta_{x',x+\hat{\mu}} U_{x\mu} - h.c.]
+ m_{0}\delta_{\mu\nu}  \label{eqn:fermion_matrix}
\end{equation}
Here, \small$\xi_{\mu}(x) = \lambda_{\mu} \eta_{\mu}(x)$\normalsize.  The anisotropies enter as $\lambda_{\mu}$:  \small $\lambda_{1}=\kappa^{- \frac{1}{2}}$, $\lambda_{2}=\kappa^{\frac{1}{2}}$, $\lambda_{3}=1$\normalsize. \small$\eta_{\mu}(x)=(-1)^{x_{1} + ... + x_{\mu}}$ \normalsize is the Kawomoto-Smit phase. \small $U_{x\mu}\equiv \exp(i\theta_{x\mu})$\normalsize, and the noncompact action is defined by 
$	\Theta_{\mu\nu}(x) = \Delta_{\mu}^{+}\theta_{\nu}(x)-  \Delta_{\nu}^{+}\theta_{\mu}(x) $.
$N=1$, which gives us $N_{f}=2$ in the continuum limit as required.  Note that this lattice action is intended to correspond to the anisotropy structure of the first flavour in (\ref{eqn:fullQED3}) with $\delta=1$.

\section{Preliminary results and comments}
\begin{figure}[t]
	\input{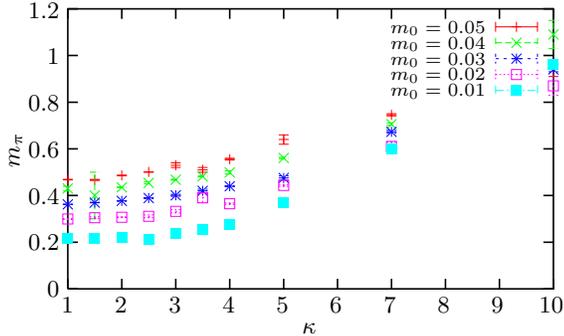}
	\vspace{-\textfloatsep}
	\caption{$m_{\pi}$ versus $\kappa$ for $m_{0}$ between 0.01 and 0.05.} \label{fig:pimass}
\vspace{-.75\textfloatsep}
\end{figure}
Here we present preliminary results concerning the chiral condensate, pion mass and renormalised anisotropy $\kappa_{r}$.  

The chiral condensate, shown in Figure \ref{fig:chircon}, exhibits a marked decrease as $\kappa$, the bare anisotropy, increases.  This could be either a phase transition at $\kappa\approx5$ or a crossover; work needs to be done regarding the susceptibilities in order to clarify this.

The pion mass, extracted from the timeslice propagator, is shown in Figure \ref{fig:pimass}.  We must caution that the error estimates are somewhat conjectural; the pions appear to be too light for the lattice.  Naively, one would expect that the pion mass would be unaffected by the change in anisotropy, as $\kappa$ does not enter (\ref{eqn:fermion_matrix}) in the $\tau$ direction.  However, as we see, the mass increases roughly quadratically with respect to $\kappa$.  This has yet to be explained; though it is plausibly due to some finite size effect, genuine physics can't be ruled out at this stage.

\begin{figure}[t]
	\input{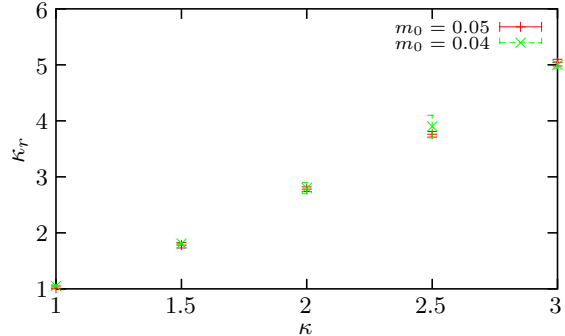}
	\vspace{-\textfloatsep}
	\caption{$\kappa_{r}$ versus $\kappa$ for $m_{0}$ between 0.04 and 0.05.} \label{fig:kappar}\vspace{-0.75\textfloatsep}
\end{figure}

Lastly, we present the renormalised anisotropy $\kappa_{r}$ (Figure \ref{fig:kappar}), which is defined by the ratio of the effective pion masses (the inverted correlation lengths) in the $x$ and $y$ directions.  \cite{kn:Herb1} present results from a large-$N$ expansion suggesting that $\kappa$ is an irrelevant parameter.  However, here we can see that $\kappa_{r}$ is greater than $\kappa$ for $1<\kappa<3$, which suggests that it is a relevant parameter.  Whether this contradicts \cite{kn:Herb1} or not is not certain --- {\em q.v.} the discussion at the end of the previous section on the anisotropy structure of our model.
As has been emphasised, the results given here are very preliminary; certain immediate research directions have been outlined in the discussion above.  It will be necessary to simulate on larger lattices with different values of $\beta$ (how $\kappa_{r}$ depends on this parameter is an intriguing question) in order to take the continuum limit, a task that has often proven difficult with QED$_{3}$.  We hope, however, that we can still shed some light on the behaviour of this novel field theory and its connection with a fascinating area of condensed matter physics.

\end{document}